\documentclass[usegraphicx,usenatbib]{mn2e}
\usepackage{natbib}

\catcode`\@=11
\def\gta{\ifmmode{\,\mathrel{\mathpalette\@versim>\,}}
    \else{$\,\mathrel{\mathpalette\@versim>}\,$}\fi}
\def\lta{\ifmmode{\,\mathrel{\mathpalette\@versim<\,}}
    \else{$\,\mathrel{\mathpalette\@versim<}\,$}\fi}
\def\@versim#1#2{\lower 2.9truept \vbox{\baselineskip 0pt \lineskip
    0.5truept \ialign{$\m@th#1\hfil##\hfil$\crcr#2\crcr\sim\crcr}}}
\catcode`\@=12  

\def\kB{k_{\rm B}}

\def\kms{\,{\rm km}\,{\rm s}^{-1}}
\def\s{\,{\rm s}}
\def\Myr{\,{\rm Myr}}
\def\sm1{\,{\rm s}^{-1}}

\def\kpc{\,{\rm kpc}}
\def\erg{\,{\rm erg}}
\def\fracj#1#2{{\textstyle{#1\over#2}}}

\begin{document}

\title[Bubbles as tracers of heat input to cooling flows]
{Bubbles as tracers of heat input to cooling flows}

\author[J. Binney, F. Alouani Bibi \& H. Omma]{J. Binney, F. Alouani
Bibi\thanks{E-mail:  binney@thphys.ox.ac.uk; alouani@astro.ox.ac.uk},  H. Omma
\\
St Johns Research Centre, 45 St Giles, Oxford, OX1 3JP, UK\\}

\date{Submitted 2006, November 21}

\pagerange{\pageref{firstpage}--\pageref{lastpage}} \pubyear{2006}

\maketitle

\label{firstpage}

\begin{abstract}
We examine the distribution of injected energy in three-dimensional,
adaptive-grid simulations of the heating of cooling flows. We show that less
than 10 percent of the injected energy goes into bubbles. Consequently, the energy input from the
nucleus is underestimated by a factor of order 6 when it is taken to be
given by $PV\gamma/(\gamma-1)$, where $P$ and $V$ are the pressure and volume of
the bubble, and $\gamma$ the ratio of principal specific heats.

\end{abstract}


\section{Introduction}

It is now widely accepted that bipolar outflows from active galactic nuclei
(AGN) largely offset the radiative losses of the hot plasma that forms
``cooling flows'' -- systems in which there is a central depression in the
temperature of the gravitationaly trapped plasma \citep{BinneyRS,Donahue06}.
Notwithstanding this concensus about the importance of AGN for cooling-flow
evolution, considerable controversy surrounds the mechanisms by which AGN
heat cooling flows. 

 Observations, first by ROSAT \citep{BohringerEtal93} and later by Chandra
\citep{FabianEtal01} of regions of depressed X-ray emission and (usually)
enhanced synchrotron emission, have played a crucial role in overturning the
paradigm of distributed mass dropout that dominated the field in the 1980s
and 1990s.  These regions have been called ``bubbles'' and considerable
effort, both numerically \citep{ChurazovEtal00,BruggenKCE02} and
theoretically \citep{Churazov02,Roychowdhury,Roychowdhuryb} has been devoted to understanding the way
bubbles interact with the surrounding, higher-density plasma.

Since lines of sight through a bubble also pass through large volumes of
X-ray emitting plasma in front of and behind a bubble, even a weak
depression in the X-ray surface brightness at the location of the bubble may
be indicative of a large drop in  emissivity inside the bubble. In fact \cite{fabian00}
report that the data for the Perseus cluster are consistent with zero emissivity inside one
of the bubbles. Consequently, it is generally assumed that inside bubbles
the density of thermal plasma is so low that the plasma emits negligibly in
X-rays. This material is generally assumed to be matter shot out from the
nucleus in a high-velocity jet, which was thermalized to a high temperature
in the hotspot that marks the interface between the jet and shocked ambient plasma.

Given that the plasma is close to an ideal gas, the thermal energy
inside a bubble is $U=PV/(\gamma-1)$, where $\gamma$ is the ratio of
principal specific heats. By taking the bubble to be in pressure balance
with the X-ray emitting plasma immediately outside it, $P$ can be determined
from the X-ray data, while $V$ follows from the geometry. Hence the energy
{\it inside\/} bubbles can be reliably estimated from observational data. 

During inflation of a bubble work is inevitably done on the surroundings plasma.
The {\it minimum\/} work that must be done at this stage is the work that
would be done if the bubble expanded reversibly, which is $PV$, bringing the
minimum energy associated with a bubble to $\gamma/(\gamma-1)PV=\fracj52PV$
for a non-relativistic plasma ($\gamma=\fracj53$). However, the
inflation of bubbles is probably anything but reversible, and in this
case, by Clausius' inequality the work done will exceed $PV$. We use
high-quality hydrodynamical simulations of bubble formation to examine how
much work is actually done by the jet, and conclude that it exceeds $\frac52PV$ by a
substantial factor.

\section{The simulations}

The framework within which we simulate cluster heating has been described
elsewhere \citep{Ommaetal,OmmaB}. We use the adaptive-grid Eulerian
hydrocode ENZO \citep{Bryan} with a piece-wise-parabolic Riemann solver. The plasma
is assumed to be an ideal, monatomic, non-relativistic  gas.
Radiative losses are calculated in the optically-thin limit using the line
and continuum cooling rates of \cite{SutherlandD}.

The initial conditions are modelled on the current configuration of the
Hydra cluster as determined from Chandra data by \cite{DavidEtal00}. Jets
are launched by depositing mass, momentum and energy in an injection disk of
radius $3\kpc$ \citep{Ommaetal}. The grid cells are cubical and $0.6\kpc$ on
a side at the centre, growing in stages by a factor of two in side length up
to $36\kpc$ at the edge of the grid, which is a cube $614\kpc$ on a side.
Periodic boundary conditions are imposed on this outer boundary. In contrast
to many other simulations of cluster heating
\citep[e.g.,][]{BassonA,Heinzetal02,VernaleoR} there is no inner boundary.

We have analyzed the energetics of all the simulations described in
\cite{OmmaB}. These differed in the amount of time the cluster cooled
passively before jet igniton, and in the power and duration of the jets, as
well as in whether there was more than one episode of jet ignition. However,
we found similar energetics of bubbles in all simulations, so here we
present only results from simulation S2, in which the cluster cooled
passively for $300\Myr$ before a pair of jets with combined power
$10^{45}\erg\s^{-1}$ ignited and ran for $25\Myr$.

\begin{figure}
\includegraphics[width=\hsize]{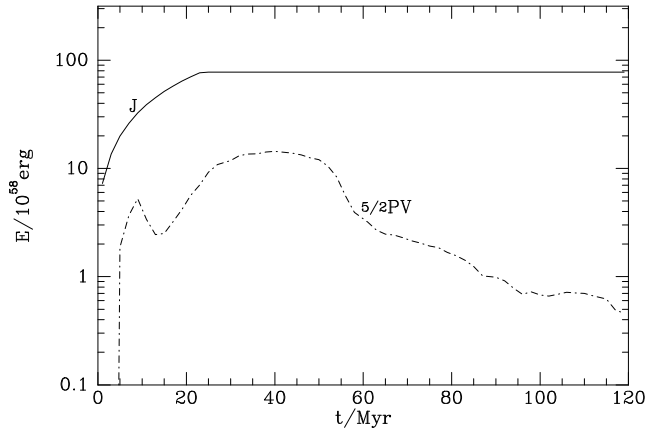}
\caption{The full curve shows the energy actually injected by the jets,
while the broken curve shows the energy associated with bubbles.\label{timefig}}
\end{figure}

\section{Results}

In a series of snapshots bubbles were identified as regions in which the
density was a factor $f$ smaller than the mean density at that radius,
and the volume $V$ of the bubbles determined as a function of $f$. Clearly
$V$ is zero for $f=0$ and is comparable to the total available volume for
$f=1$.  For $0.2\lta f\lta 0.75$ there is a plateau in $V$ that we
identify with the bubbles, and we present results obtained by adopting
$f=0.3$.

\begin{figure*}
\includegraphics[width=.8\hsize]{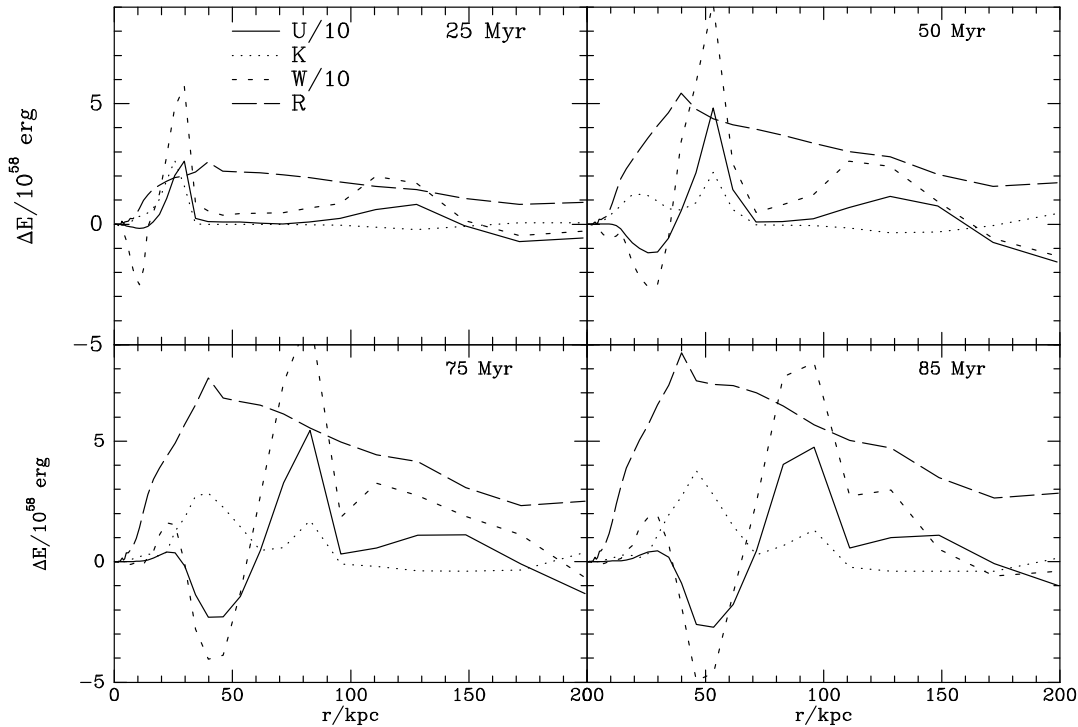}
\caption{The radial distribution of energy changes defined by equation (\ref{eqone}) since jet ignition. \label{radfig}}
\end{figure*}

In Fig.~\ref{timefig} the full curve shows as a function of time since jet
ignition the energy injected into the simulation, while the dashed curve
shows $\fracj52PV$ for the bubbles.  The energy associated with the bubbles
peaks at $14\times10^{58}\erg$, while the energy injected is
$78\times10^{58}\erg$. Thus less than a fifth of the injected energy is ever
associated with identifiable bubbles, and at most epochs the fraction is
less.

To answer the question ``where {\em is\/} the energy?'' we have analysed
the volume within $200\kpc$ of the cluster centre in the following way. We
grouped the computational cells into  approximately 40 spherical shells and in
each shell $i$ calculated
 \begin{equation}\label{eqone}
\Delta E_i=\sum_{\rm cells\ \alpha}V_\alpha[e_\alpha(t)-e_\alpha(0)],
\end{equation}
 where $V_\alpha$ is the volume of the $\alpha$th cell and $e_\alpha$ is an
energy density within that cell. The energy densities considered were
$u=\fracj32n\kB T$, where $n$ is the particle density, $w=-\rho \Phi$, where
$\rho$ is the matter density and $\Phi<0$ is the gravitational potential,
and $k=\fracj12\rho v^2$, where $v$ is the flow speed. In this way we obtain
measures $\Delta U$, $\Delta W$ and $\Delta K$ of the changes since jet
ignition in internal, potential and kinetic energy within each shell.

Fig.~\ref{radfig} gives the radial distribution of energy differences at
four times, $25$, $50$, $75$ and $85\Myr$ after jet ignition. At $25\Myr$
(top left panel) the jets have just switched off and the curves are
structured by two distinct processes. At $r\lta30\kpc$ the jets are
dominant, producing sharp peaks in internal (full curve), potential (short-dashed
curve), and kinetic energy (dotted). Interior to the peak in potential energy
$\Delta W$ at $r\simeq20\kpc$, there is a region in which $\Delta W<0$
because uplift has reduced the mass density there. In this region the
internal energy is only very slightly depressed because the drop in $\rho$
is compensated by a rise in $T$. Beyond $\sim40\kpc$ the curves for $\Delta
U$ and $\Delta W$ are structured by radiative cooling, which causes a slow
decline in $T$ and a flow of material from large to small radius that causes
$\Delta W$ to be positive at $r\lta150\kpc$ and negative further out.

\begin{table}
\begin{center}
\begin{tabular}{l|rrrr|rrrr}
$t/\!\Myr$&$\Delta U$&$\Delta K$&$\Delta W$&$\Delta E$&$\Delta U$&$\Delta K$&$\Delta W$&$\Delta E$\\
\hline
 25&  -11&    2& -132&  123&   64&    8&  160&  -88\\
 50&  -54&    9& -143&   98&   90&    4&  215& -120\\
 75&  -62&   15&  -82&   34&   95&    3&  223& -125\\
 85&  -63&   16&  -60&   13&   98&    2&  211& -111\\
\end{tabular}
\end{center}
\caption{Energy changes (in units of $10^{58}\erg$) in the radius ranges
$(0,r_{\rm cavity})$ (columns 2 to 5) and
$(r_{\rm cavity},r_{\rm shock})$ (columns 6 to 9).
The values of $r_{\rm cavity}$ are 15, 37, 58 and $70\kpc$, while those of
$r_{\rm shock}$ are 35, 75, 100 and $115\kpc$. $\Delta E\equiv\Delta
U+\Delta K-\Delta W$.\label{etable}}
\end{table}

The long-dashed curve in the top left panel shows the radiated energy $R$:
unlike the other curves, this is not calculated by differencing energy
densities at two epochs but by integrating the X-ray emissivity $j_X$
through the volume of each cell and with respect to time since jet ignition.

The top right panel of Fig.~\ref{radfig} shows the disposition of energy
$50\Myr$ after jet ignition. Compared to the top left panel for $t=25\Myr$,
the long-dashed curve of the radiative losses has risen at all radii, and
the peaks in $\Delta U$, $\Delta W$ and $\Delta K$ have moved outwards with
the rising jet debris. Now the curves for $\Delta U$, $\Delta W$ and $\Delta
K$ are structured by the jet for $r\lta70\kpc$ and by cooling further out.
At small radii the peak in $\Delta U$ is now preceeded by a pronounced
trough, reflecting the development of an under-pressured region behind the
zone filled by material that is being carried up by its momentum, and
decelerating in the cluster's gravitational field. At $r\gta80\kpc$ cooling
continues to drive an inward flow that makes the negativity of $\Delta U$
and $\Delta W$ at large radii more pronounced.

The bottom left panel of Fig.~\ref{radfig}, for $t=75\Myr$ after jet
ignition, shows a new feature: the curves for $\Delta U$ and especially
$\Delta W$ have developed peaks at $r\simeq20\kpc$, interior to the troughs
that preceed the main peaks. This new region of enhanced potential energy
reflects material that has fallen down after being uplifted and is now at a
radius lower than the radius it would have in an equilibrium model, given
its specific entropy. Thus this peak reflects a particular phase of the
gravity waves that are strongly excited by the jets. The bottom-right panel,
for $r=85\Myr$ is very similar to that for $75\Myr$.

At each panel of Fig.~\ref{radfig} one can identify two characteristic
radii, namely the radius $r_{\rm shock}$ that marks the end of the sharp
drop in curves for $\Delta U$ and $\Delta W$ that the shock produces, and
the radius $r_{\rm cavity}$ at which these curves cross zero on the other
side of the peak.  For the time of each panel in Fig.~\ref{radfig}, Table
\ref{etable} gives energies, in units of $10^{58}\erg$, contained in the
regions $r<r_{\rm cavity}$ and $r_{\rm cavity}<r<r_{\rm shock}$. We see that
at $25\Myr$, as the jet shuts off, the total energy inside $r_{\rm cavity}$
has increased by an amount $\Delta E=\Delta U+\Delta K-\Delta
W=123\times10^{58}\erg$, that greatly exceeds the total injected energy,
$E_{\rm jet}=78\times10^{58}\erg$ and is more than an order of magnitude
greater than the energy then associated with bubbles ($8.9\times10^{58}\erg$
from Fig.~\ref{timefig}). This energy increase is largely caused by a
negative value of $\Delta W=-132\times10^{58}\erg$ following uplift of
material -- to achieve this value, the jet has not in fact supplied
$132\times10^{58}\erg$ of energy, which would suffice to lift the uplifted material to
$r=\infty$, but only the smaller amount required to move material out of the
cavity zone. This explains  why $\Delta E$ for this region exceeds the the jet
energy. In the dense region outside $r_{\rm cavity}$ and inside $r_{\rm
shock}$, $\Delta W$ has increased by twice the jet energy, reflecting (i)
the arrival from below of gravitationally bound uplifted material, and (ii)
the arrival from above of material that has cooled. The internal energy of the
gas in this region has risen by 75\% of $E_{\rm jet}$. At this stage the bulk kinetic
energy alone is equal to the energy associated with bubbles.

At later times we see that within $r_{\rm cavity}$, $\Delta U$ and $\Delta K$ continue to
increase, while $\Delta W$ becomes less negative. In the dense shell further
out $\Delta K$ continually decreases, while $\Delta U$ slowly rises. The
total energy change $\Delta E$ in the dense shell
is always falling as material is driven in by the jet-induced flow  on one
side and gravity on the other, bringing with it large (negative) potential
energies. In this region cooling also attenuates the rise in $\Delta U$ that
would otherwise occur.

\section{Discussion}

We have shown that in the simulations
the minimum energy
associated with bubbles, $E_{\rm min}=\fracj52PV$, is smaller than the injected
energy by at least a factor of nearly 6. Why is this so, and can we expect the
result to extend to real cooling flows, or does it merely reflect
limitations of the simulations? 

The factor $\fracj52$ allows for the energy $\frac32PV$ that is actually in
the bubble, and an allowance $PV$ for the work that must be done to create a
cavity in the ambient medium. As explained in the Introduction, this is the
{\em minimum\/} allowance, and is the actual requirement only if the cavity
is created reversibly and at constant pressure. In this case the displaced
material would be neither heated nor accelerated, and the work $PV$ would be
done on the mechanism (piston etc) that held the pressure constant. In
reality the creation of the cavity is strongly irreversible, so the
displaced material is heated and accelerated. In as much as real jets
probably involve larger velocities than those ($\sim40\,000\kms$)
characteristic of the simulations, the simulations may be expected to under-
rather than over-estimate the irreversibility of bubble inflation.

The internal energy of displaced material is changed by of order itself, and
its original value was equal to the bubble energy. Moreover the material
evacuated from the cavity itself displaces material, so the mass of the
material that has to be moved and heated exceeds the mass originally
associated with the cavity by a factor $\alpha>1$, say. Doubling the
internal energy of the shifted material, would involve supplying
$\frac32\alpha PV$ of energy over and above what is required in the
reversible case. 

In the simulations the timescale for bubble inflation is of order the
sound-crossing time at the ambient sound speed, and it is widely assumed
that the same is true in real clusters. Consequently, the shifted material
inevitably gains an amount of kinetic energy that is comparable to the
original internal energy, $\fracj32\alpha PV$, as Table \ref{etable}
confirms to be the case. 

These two estimates of energy that must be supplied should be added to the
energy $\gamma PV/(\gamma-1)$ required in the reversible case, where
$\gamma$ is the ratio of principal specific heats of the material in the
cavity.  In our simulations $\gamma=\frac53$, so the actual energy required
is larger than the minimum energy by a factor $1+\fracj65\alpha$. If the
cavity contains relativistic plasma, as is frequently
assumed, and the ambient plasma is non-relativistic,
the energy that must be supplied is larger than the minimum energy
by a factor $1+\frac34\alpha$.  The energetics of the simulations implies
that the shifted mass is $\alpha=3.75$ times the mass originally in the
bubble 

In reality $\alpha$ is probably not so large because diffusivity will lead
to $PV$ being underestimated, both in the simulations and in real cooling
flows.  Our simulations do not explicitly include thermal conductivity, but
any numerical hydrodynamic scheme has significant diffusivity on the scale
of the grid (or the mean-free path in the case of smooth-particle
hydrodynamics) by virtue of assigning single values to the dynamical
variables at each grid point, when in reality the material in a cell would
contain hot and cold regions. As a consequence, energy can cross a cell in
one timestep when hot matter enters on one side at the start of a timestep,
and warmed matter leaves on the opposite side at the start of the next
timestep. There is no concensus regarding the appropriate values of the
kinetic transport coefficients of intracluster gas [see \cite{KimN,Pope} for
conflicting positions], and even less understanding of the effectiveness of
turbulent transport.  However, turbulence stretches and distorts bubbles,
and when their narrowest dimension becomes comparable to the grid spacing,
diffusivity will eliminate them. Thus kinetic and turbulent diffusivity
cause bubbles to shrink on a dynamical timescale in the
simulations, as they probably do in reality

\begin{figure}
\includegraphics[width=\hsize]{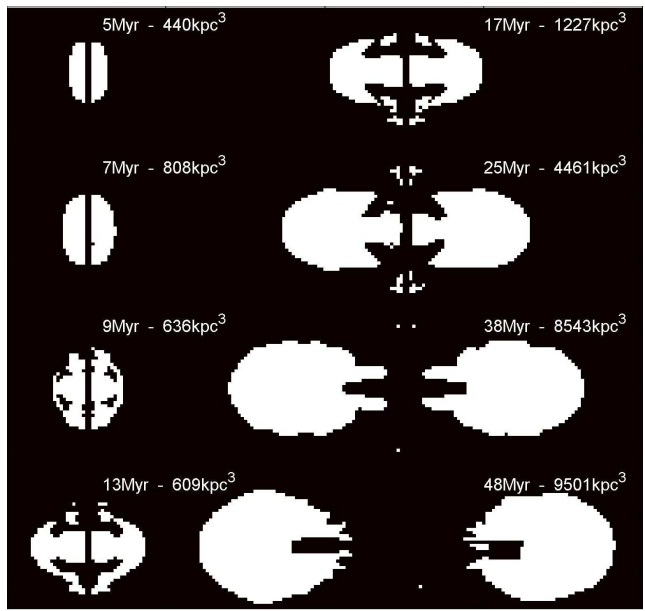}
\caption{Cross sections through the cavities at eight times. In addition to
the time since jet ignition, each panel gives the total volume of the
cavities. Note that there are scale changes between images.\label{cavfig}} 
\end{figure}

The broken curve in Fig.~\ref{timefig} illustrates this phenomenon at two
points: between $t\simeq8$ and $14\Myr$, $PV$ dips while the jet is still
firing, and then at $t>50\Myr$ it falls continuously after the jet has
switched off. The dip at early times reflects shrinkage in the volume of
low-density material by $\sim25$ percent following the emergence of
turbulent eddies around the jet; as the jet establishes itself at
$t\approx7\Myr$, the cavities shift from being nearly hemispherical to
elongated (Fig.~\ref{cavfig}). The steep velocity gradient perpendicular to
the jet axis then induces large-scale turbulence, which draws cool material
into the cavity.  The ensuing fragmentation of the cavity leads to the decline in
cavity volume that causes the dip in the $PV$ curve of Fig.~\ref{timefig}.
At $t>14\Myr$ a new regime is entered in which lobes swell around the now
more slowly advancing hotspots. The drop in the $PV$ curve at $t>50\Myr$ is
also driven by turbulent breakup. How realistic the steepness of this drop
is, we do not know.

Observationally it is inevitable that the volume of cavities will be
underestimated because it has been determined from X-ray surface brightness,
rather than from a knowledge of the plasma density at each point in three
dimensions, as can be done in the simulations. Clearly such underestimation
of the volume will lead to a corresponding underestimation of the power
input by the jets. It would be interesting to compare estimates of cavity
volumes obtained from X-rays with ones obtained from high-resolution radio
maps, which might be less sensitive to projection effects. On the other hand,
the X-ray cavities sometimes do not coincide with the radio lobes -- the
Hydra cluster provides an example of this phenomenon \citep{Nulsen}.

This discussion leads to the conclusion that taking the energy injected by
jets to be $\fracj52PV$ probably does lead to its being underestimated by a
substantial factor, which we have estimated to be $\sim5.5$. Our estimate of
this factor may be on the high side if the simulations over-estimate the
diffusivity of the plasma that arises from both turbulent mixing and kinetic
processes, but may be on the low side because our jet speeds are relatively low
and cavity volumes are likely to be underestimated from X-ray surface
brightnesses.

The fact that most of the injected energy is never in cavities, has
important implications for the time- and space-dependence of cluster
heating. As Table \ref{etable} shows, much of this energy is
stored as potential energy -- the energy required to lift low-entropy gas
above its natural position in the cluster atmosphere. This energy
subsequently heats the plasma through the dissipation of the non-linear
gravity waves to which it gives rise. The timescale of this release can be
very long, especially in the case of an unusually violent outburst of the
AGN, which drives cool plasma to great heights, at which the dynamical time
can be long \citep{Ommaetal}.

Several studies \citep[e.g.,][]{Birzan,NipotiB,Bestetal,Allenetal06} have
considered the implications of cooling flows for the phenomenolgy of AGN and
radio sources under the assumption that the injected energy is $\gamma
PV/(\gamma-1)$.  The studies conclude that heating by AGN can balance
radiative cooling provided the duty cycle of the jets is high. With our
increased estimate of the energy injected in a given episode of bubble
inflation, the required duty cycles will be shorter.

Assuming a 10 percent efficiency in the conversion of rest mass to jet
power, \cite{Allenetal06} conclude that jet powers are smaller than the rate
of energy release from Bondi accretion of hot plasma by a factor $\sim 6$
(their Fig.~4). Prima facie, our increase in the required jet power would
bring it up to the rate of release of energy by Bondi accretion. Is this
conclusion problematic, given that only a fraction of the plasma that passes
over the Bondi accretion radius can be swallowed by the black hole?

The rate at which material passes the Bondi radius provides a useful measure
of the mean jet power, but not of the instantaneous power, because inside
the Bondi radius an accretion disc is likely to provide a reservoir of
material, and jet outbursts are thought to be brief episodes in the evolution
of this disc \citep{NipotiBB,Koerding}. Thus on the timescale of bubble
inflation, jet power can safely exceed the mean power available by Bondi
accretion, even if  the black hole's only mass supply is from
virial-temperature gas (which is by no means certain). Estimates of the
power available from Bondi accretion are most usefully compared with the
radiative losses of the cooling flow \citep{BinneyRS}.

\section*{Acknowledgments}

This work has been supported by St Johns College, Oxford

\label{lastpage}

\end{document}